\documentclass[prl,10pt,twocolumn,superscriptaddress,floatfix,showpacs]{revtex4-1}

\usepackage[utf8]{inputenc}  
\usepackage[T1]{fontenc}     
\usepackage[british]{babel}  
\usepackage[sc,osf]{mathpazo}
\usepackage[scaled=0.86]{berasans}  
\usepackage[colorlinks=true, citecolor=blue, urlcolor=blue]{hyperref}  
\usepackage{graphicx} 
\usepackage[babel]{microtype}  
\usepackage{amsmath,amssymb,amsthm,bm,amsfonts,mathrsfs,bbm} 

\usepackage{xspace}  
\usepackage{pgfplots}
\usepackage{xcolor,colortbl}

\begin{document}

\title{Nonlocal correlations: Fair and Unfair Strategies in Bayesian Game}


\author{Arup Roy}
\affiliation{Physics and Applied Mathematics Unit, Indian Statistical Institute, 203 B. T. Road, Kolkata 700108, India.}

\author{Amit Mukherjee}
\affiliation{Physics and Applied Mathematics Unit, Indian Statistical Institute, 203 B. T. Road, Kolkata 700108, India.}

\author{Tamal Guha}
\affiliation{Physics and Applied Mathematics Unit, Indian Statistical Institute, 203 B. T. Road, Kolkata 700108, India.}

\author{Sibasish Ghosh}
\affiliation{Optics \& Quantum Information Group, The Institute of Mathematical Sciences, C.I.T Campus, Tharamani, Chennai 600 113, India.}

\author{Some Sankar Bhattacharya}
\affiliation{Physics and Applied Mathematics Unit, Indian Statistical Institute, 203 B. T. Road, Kolkata 700108, India.}

\author{Manik Banik}
\affiliation{Optics \& Quantum Information Group, The Institute of Mathematical Sciences, C.I.T Campus, Tharamani, Chennai 600 113, India.}


\begin{abstract}
Interesting connection has been established between two apparently unrelated concepts, namely, quantum nonlocality and Bayesian game theory. It has been shown that nonlocal correlations in the form of advice can outperform classical equilibrium strategies in common interest Bayesian games and also in conflicting interest games. However, classical equilibrium strategies can be of two types, fair and unfair. Whereas in fair equilibrium payoffs of different players are same, in unfair case they differ. Advantage of nonlocal correlation has been demonstrated over fair strategies. In this work we show that quantum strategies can outperform even the unfair classical equilibrium strategies. For this purpose we consider a class of two players games which as a special case includes the conflicting game proposed in [\href{http://journals.aps.org/prl/abstract/10.1103/
PhysRevLett.114.020401}{Phys. Rev. Lett. 114, 020401 (2015)}]. These games can have both fair and unfair classical equilibria and also can have only unfair ones. 
We provide a simple analytic characterization of the nonlocal correlations that are advantageous over the classical equilibrium strategies in these games.   
\end{abstract}


\maketitle

	
Undoubtedly one of the most fundamental contradictions of Quantum mechanics (QM) with classical physics gets manifested in its nonlocal behavior. This bizarre feature of QM was first established in the seminal work of J. S. Bell \cite{Bell64}, where he has shown that QM is incompatible with the \emph{local-realistic} world view of classical physics. More precisely, Bell showed that measurement statistics of multipartite entangled quantum systems can violate an empirically testable local realistic inequality (in general called Bell type inequalities) which establishes the denial of \emph{local realism} underlying QM. Since Bell's work, nonlocality remains at the center of quantum foundational research (see \cite{Reviw} and references therein) and it has been verified in numerous successful experiments, starting from the famous Aspect's experiment \cite{Aspect81} to very recent loop-hole free tests \cite{loopfree}. Apart from foundational interest quantum nonlocal correlations have been proved to be the key resource in various device-independent protocols \cite{Applications}. Very recently Brunner and Linden have established that Bell nonlocality has interesting connection with a seemingly  different area of research, namely, theory of Bayesian game \cite{Brunner13}. A Bayesian game can be played under classical equilibrium strategies which are of two types, fair equilibria and unfair equilibria. Whereas in fair equilibria payoffs of different parties are same, in unfair equilibria they differ. It has been shown that QM can provide advantageous strategies over the best classical strategies in common interest Bayesian games \cite{Brunner13} and can also outperform the fair classical equilibrium strategies in conflicting games \cite{Pappa15}. The aim of this present paper is to study whether nonlocal correlation can be advantageous over the classical unfair equilibrium strategies in such a games.      

Operationally Bell type inequalities can be best understood in terms of games involving several number of spatially separated parties. Each party receives some inputs and produces some outputs. No communication is allowed among the parties during the game but, they can share correlations. The aim is to cooperatively optimize some payoff function. If the correlation shared among them is classical one (or more precisely to say \emph{local realistic} one) then the payoff is upper bounded by some threshold value. However it may be possible that using correlation of entangled quantum states these optimal classical bounds can be superseded, which establishes nonlocal behavior of the shared entangled state. The canonical example of such game is Clauser-Horne-Shimony-Holt (CHSH) game \cite{Clauser69} involving two parties, say Alice and Bob, each with binary inputs and binary outputs. Whereas payoff of this game in local realistic theory is upper bounded by $3/4$, in quantum theory one can achieve it up to $1/2(1+1/\sqrt{2})$. 

Game theoretic formulation of Bell type inequalities prompted Brunner \emph{et al.} to explore the connection between Bell nonlocality and Bayesian game theory \cite{Brunner13}. The theory of Bayesian game was discovered by Harsanyi and the framework is developed for games with incomplete information  \cite{Harsanyi67} (see also \cite{Osborne03}). In such a game, each player may have some private information unknown to other players; on the other hand the players may have a common piece of advice and thus can follow correlated strategies, giving rise to the concept of correlated Nash equilibrium \cite{Aumann74}. As pointed out in \cite{Brunner13}, the concept of private information in Bayesian games is analogous to the notion of locality in Bell inequalities (BI). And the fact that common advice in Bayesian games does not reveal the private information mimics the concept of no-signaling resources in case of BI. A larger class of such no-signaling correlations, proven to be stronger resource than local realistic correlations in nonlocal games, are available in QM. Naturally the question arises whether such nonlocal quantum correlations indeed provide advantage in Bayesian games over the classical strategies. Interestingly, in their paper Brunner \emph{et al}. have answered affirmatively to this question \cite{Brunner13}. The CHSH game as well as GHZ game \cite{Greenberger90}, Mermin game \cite{Mermin90a}, Magic Square Game \cite{Mermin90b,Peres90}, Hidden Matching game \cite{Kerenidis04, Buhrman11} and the three games of \cite{Brunner13} are all example of common interest games where the involving parties have to optimize some payoff, cooperatively. On the other hand in conflicting interest game interests of the players differ, resulting to conflict in their best actions. Battle of Sexes (BoS) is a classic example of such conflicting interest game. Interestingly in \cite{Pappa15}, the authors have provided an example of a conflicting game where quantum strategies can outperform the classical fair equilibrium strategies. Furthermore, using semi-definite programming (SDP) they have shown that Bell states along with suitably chosen measurements are the quantum equilibrium strategies for the said game. Using the initials of the authors let call this PKLSZDK game. 

A Baysian game can have only fair equilibria, both fair and unfair equilibria or only unfair equilibria. In this present paper we address the question whether quantum advice can surpass the classical unfair equilibrium strategies in such games. Interestingly, we answer affirmatively to this questions. For this purpose we have constructed a two parametric class of two players game, each player having two types/inputs and two possible actions/outputs for each type. As a special case, the PKLSZDK game is also included in this class. For certain conditions on the parameters we show that these games can have only unfair equilibria. To show the advantages of nonlocal correlation over the classical equilibrium strategies we take a very analytic approach. First we show that in the $2-2-2$ scenario (i.e. two-party, each with two measurements and each measurement with two outcomes)  any no signaling correlation can be expressed in a canonical form. Using this canonical form we completely characterize the no-signaling correlations providing advantages over the classical fair and unfair strategies in these games. One interesting consequence of our analysis is that it provides a simple explanation (without using any SDP), why the said strategies of \cite{Pappa15} are quantum fair equilibrium one. 
  
\emph{A class of two players Bayesian games}: Here we adopt the same notations used in Ref.\cite{Pappa15}. Let Alice and Bob are two players involved in the game. Alice's and Bob's types/inputs are denoted as $x_A\in\mathcal{X}_A$ and $x_B\in\mathcal{X}_B$, respectively. For each types they take some actions/outputs  denoted as $y_A\in\mathcal{Y}_A$ and $y_B\in\mathcal{Y}_B$ and accordingly they are given payoffs/utilities denoted as $u_A$ and $u_B$, respectively, where $u_i:\mathcal{X}_A\times\mathcal{X}_B\times
\mathcal{Y}_A\times\mathcal{Y}_B\rightarrow \mathbb{R}$    , for $i\in\{A,B\}$. For the class of games considered here, $\mathcal{X}_A=\mathcal{X}_B=
\mathcal{Y}_A=\mathcal{Y}_B=\{0,1\}$ and the utilities are given in Table-\ref{table1}. In accordance with the parameter $\kappa$ and $\tau$ of Table-\ref{table1} let us denote such a game as $\mathcal{G}(\kappa,\tau)$. Whenever $\kappa< \tau$, there is a conflict between Alice and Bob in choosing their actions. Note that the game $\mathcal{G}(1/2,1)$ is the conflicting game studied in \cite{Pappa15}. 
\begin{center}
\begin{table}[t!]
\caption{Utility table for the game $\mathcal{G}(\kappa,\tau)$. Both $\kappa$ and $\tau$ are positive.}
\begin{tabular}{ p{1.2cm} | p{3.3cm} || p{3.3cm} }
  \hline\hline			
   & ~~~~~~$x_A\wedge x_B=0$ & ~~~~~~~~~$x_A\wedge x_B=1$ \\
  \hline\hline
   & ~$y_B=0$~~~~~~$y_B=1$ & ~~~$y_B=0$~~~~~~$y_B=1$ \\
  \hline
  $y_A=0$ & $(1,\kappa)$ ~~~~~~$(0,0)$&~~~~~$(0,0)$ ~~~$(3/4,3/4)$ \\
  $y_A=1$ &~ $(0,0)$ ~~~~~$(1/2,\tau)$&~$(3/4,3/4)$ ~~$(0,0)$ \\
  \hline  
\end{tabular}
\label{table1}
\end{table}
\end{center} 
 \vspace{-.7cm}
 
In the case of correlated strategies, i.e., when the parties are given some common advice, the average payoff is calculated as:
\begin{equation}	
F_i=\sum_{x,y} P(x)P(y|x) u_i(x,y).\label{ap}
\end{equation}				
Here $P(x)$ is the probability distribution over the Alice's and Bob's joint type $x\equiv(x_A,x_B)$ which is considered to be uniform for this particular game and $P(y|x)$ denote the conditional joint action $y\equiv(y_A,y_B)$ given the type $x$, i.e., the probability that Alice takes action $y_A$ and Bob takes action $y_B$ given their joint type $(x_A,x_B)$. In the case of such correlated strategies, Aumann introduced the concept of correlated equilibria (CE) \cite{Aumann74}, which have several nice properties: they are easier to find \cite{Papadimitriou08}, every Nash equilibrium is a CE and convex combinations of CE are again CE. For playing the game $\mathcal{G}(\kappa,\tau)$ each of Alice and Bob can take one of the following four pure classical strategies:
\begin{equation}
g^1_i(x_i)=0;~g^2_i(x_i)=1;~g^3_i(x_i)=x_i;
~g^4_i(x_i)=x_i\oplus 1;\nonumber \label{cs}
\end{equation}
where $g^1_i(x_i)=0$ means that $i^{th}$ party takes the action $0$ whatever be the type is and similarly the other cases; $\oplus$ denotes modulo $2$ sum. For $16$ possible pure strategies of Alice and Bob (together) their average payoffs $(F_A,F_B)$ are listed in Table-\ref{table2}. As discussed earlier, $\mathcal{G}(\kappa,\tau)$ be a conflicting interest game when $\tau>\kappa$. In this case when $\kappa<\frac{3}{4}$ the strategies $(g_A^1,g_B^3)$, $(g_A^3,g_B^4)$, and $(g_A^4,g_B^2)$ are Nash equilibria and let's denote the corresponding payoffs, $(F_A^{eq_1},F_B^{eq_1})\equiv\left(\frac{11}{16},\frac{3}{16}+\frac{\kappa}{2}\right)$, $(F_A^{eq_2},F_B^{eq_2})\equiv\left(\frac{9}{16},\frac{3}{16}+\frac{\kappa+\tau}{4}\right)$, and $(F_A^{eq_3},F_B^{eq_3})\equiv\left(\frac{7}{16},\frac{3}{16}+\frac{\tau}{2}\right)$, respectively; and for $\kappa>\frac{3}{4}$ the strategies $(g_A^1,g_B^1)$, $(g_A^3,g_B^4)$, and $(g_A^4,g_B^2)$ are Nash equilibria with corresponding payoffs as $(F_A^{eq'_1},F_B^{eq'_1})\equiv\left(\frac{3}{4},\frac{3\kappa}{4}\right)$, $(F_A^{eq_2},F_B^{eq_2})\equiv\left(\frac{9}{16},\frac{3}{16}+\frac{\kappa+\tau}{4}\right)$, and $(F_A^{eq_3},F_B^{eq_3})\equiv\left(\frac{7}{16},\frac{3}{16}+\frac{\tau}{2}\right)$, respectively. For the parameter value $\tau>\kappa>1$, all three equilibrium are unfair and in every case Bob's payoff is greater than Alice's. Note that in these cases even no fair correlated equilibrium strategy is possible. The cases where $\kappa+\tau=1.5$ give a fair equilibria strategy as occurred in the conflicting game of \cite{Pappa15}. When $\kappa>\tau$ it turns out to be a game with only one equilibrium; for $\kappa<\frac{3}{4}$ the strategy $(g_A^1,g_B^3)$ is Nash equilibrium and the corresponding payoff, $(F_A^{eq_1},F_B^{eq_1})\equiv\left(\frac{11}{16},\frac{3}{16}+\frac{\kappa}{2}\right)$, and for $\kappa>\frac{3}{4}$ the Nash equilibrium is the strategy $(g_A^1,g_B^1)$ with $(F_A^{eq'_1},F_B^{eq'_1})\equiv\left(\frac{3}{4},\frac{3\kappa}{4}\right)$. Since any classical (local realistic) advice can be written as $P(y_A,y_B|x_A,x_b)=\int d\lambda P(y_A|x_A,\lambda)P(y_B|x_B,\lambda)$, with  
$\lambda$ being a local variable (also called hidden variable by the quantum foundation community), so convexity ensures that using any such advice it is not possible to overcome the equilibrium payoffs. 
However in quantum world there are no-signaling correlations that are not in this local realistic form (thus called nonlocal) and hence there may be a possibility to overcome the classical equilibrium payoffs.

\begin{center}
\begin{table}[t!]
\begin{tabular}{ l | c | c | c | r }
				\hline\hline
				& $g^1_B$ & $g^2_B$ & $g^3_B$ &$g^4_B~~~~~~~~~$\\ \hline \hline
				$g^1_A$ & \cellcolor{green!15}$\left(\frac{3}{4},\frac{3\kappa}{4}\right)$ & $\left(\frac{3}{16},\frac{3}{16}\right)$ & \cellcolor{red!25}$\left(\frac{11}{16},\frac{3}{16}+\frac{\kappa}{2}\right)$ & $\left(\frac{1}{4},\frac{\kappa}{4}\right)~~~~~~$\\ \hline
				$g^2_A$ & $\left(\frac{3}{16},\frac{3}{16}\right)$ & $\left(\frac{3}{8},\frac{3\tau}{4}\right)$ & $\left(\frac{1}{8},\frac{\tau}{4}\right)$ & $\left(\frac{7}{16},\frac{3}{16}+\frac{\tau}{2}\right)~~$\\ \hline
				$g^3_A$ & $\left(\frac{11}{16},\frac{3}{16}+\frac{\kappa}{2}\right)$ & $\left(\frac{1}{8},\frac{\tau}{4}\right)$ & $\left(\frac{1}{4},\frac{\kappa}{4}\right)$ & \cellcolor{blue!15}$\left(\frac{9}{16},\frac{3}{16}+\frac{\kappa+\tau}{4}\right)$\\ \hline
				$g^4_A$ & $\left(\frac{1}{4},\frac{\kappa}{4}\right)$ & \cellcolor{blue!15}$\left(\frac{7}{16},\frac{3}{16}+\frac{\tau}{2}\right)$ & $\left(\frac{9}{16},\frac{3}{16}+\frac{\kappa+\tau}{4}\right)$ & $\left(\frac{1}{8},\frac{\tau}{4}\right)~~~~~~$\\
				\hline
\end{tabular}
\caption{(Color on-line) Alice's and Bob's average payoffs $(F_A,F_B)$ for different pure strategies in the game $\mathcal{G}(\kappa,\tau)$. For conflicting interest case ($\tau>\kappa$) there are three equilibria when $\kappa<\frac{3}{4}$ (red and blue) and when $\kappa>\frac{3}{4}$ (green and blue), and for $\tau<\kappa$ there is one equilibrium: red when $\kappa<\frac{3}{4}$ and green when $\kappa>\frac{3}{4}$.}
\label{table2}
\end{table}
\end{center}   
\vspace{-1cm}
\emph{$2-2-2$ no-signaling correlations}: For the two-party scenario with two two-outcome measurements for each party let denote the joint probability distribution as $P(ab|ij)$, where the outcomes $a,b\in\{+,-\}$ and the measurement settings $i,j\in\{0,1\}$. We can express the joint distribution as \cite{Gazi13}:
\begin{equation}
P(ab|ij)\equiv(c_{ij},m_{ij}-c_{ij},n_{ij}-c_{ij},1-n_{ij}-m_{ij}+c_{ij}),
\end{equation}
with outcomes in the order $(++,+-,-+,--)$. Here $m_{ij}$ and $n_{ij}$ denote the corresponding marginal probabilities of Alice and Bob, with positivity imposing the restrictions, $\max\{0,m_{ij}+n_{ij}-1\}\le c_{ij}\le\min\{m_{ij},n_{ij}\}~\forall~ij$.
According to no-signaling Alice's marginal outcome probability should not depend on Bob's measurement settings and vice versa, which can be expressed as 	$m_{00}=m_{01}:=m_{0},~m_{10}=m_{11}:=m_{1},~
n_{00}=n_{10}:=n_{0},~n_{01}=n_{11}:=n_{1}$.
The celebrated Bell-CHSH expression is given by,$\mathbb{B}=\langle00\rangle+\langle01\rangle
+\langle10\rangle-\langle11\rangle$, 
where $\langle ij\rangle:=P(++|ij)-P(+-|ij)-P(-+|ij)+P(--|ij)$. A no-signaling probability distribution has a local realistic description if and only if it satisfies the Bell-CHSH inequality, i.e., \emph{iff} $|\mathbb{B}|\le 2$ \cite{Fine82}. In terms of probabilities, the Bell-CHSH expression becomes,
\begin{equation}
\mathbb{B}=2+4(c_{00}+c_{01}+c_{10}-c_{11})-4(m_0+n_0).\label{bell}
\end{equation}
Collection of all such no-signaling correlations $\mathcal{NS}$ form a $8$ dimensional polytope with $24$ vertices and the facets are determined by positivity and no-signaling constraints. Collection of all classical (local realistic) correlations $\mathcal{L}$ and all quantum correlations $\mathcal{Q}$ are convex sets lying within $\mathcal{NS}$. While $\mathcal{L}$ is again a polytope with trivial facets given by positivity and the nontrivial facets determined by Bell-CHSH inequalities, the set $\mathcal{Q}$ is convex but not a polytope \cite{Barrett05}. Any correlation outside $\mathcal{L}$ is called nonlocal and the strict set inclusion relations $\mathcal{L}\subset\mathcal{Q}\subset\mathcal{NS}$ reflect the fact that quantum theory contains nonlocal correlations, but nonlocality of quantum theory is restricted compared to general no-signaling correlations. While $\mathbb{B}$ can be at most  $2\sqrt{2}$ in $\mathcal{Q}$ \cite{Tsirelson80}, in $\mathcal{NS}$ it can go up to $4$ \cite{Popescu94}.

\emph{Nonlocal advantage over classical strategies}: In the Bayesian game described above, the two players can be commonly advised by a general no-signaling correlation. Given such an advice $NS\in\mathcal{NS}$ Alice's and Bob's average payoffs, respectively, reads:
\begin{eqnarray}\label{fa}
F^{NS}_A&=&\frac{1}{16}\left[3+3/2\mathbb{B}+2(m_0+n_0)+(m_1+n_1)
\right],\\\label{fb}
F^{NS}_B&=&\frac{1}{16}\left[(10\tau-2\kappa)+(\tau+\kappa)\mathbb{B}
+4(\kappa-\tau)(m_0+n_0)\right.\nonumber\\
&&\left. +(3-4\tau)(m_1+n_1)+4\left(\kappa+\tau-3/2\right)c_{11}\right].~~~~~
\end{eqnarray} 
A no-signaling nonlocal advice outperforms some classical equilibrium payoff $(F^{eq}_A,F^{eq}_B)$ if $F^{NS}_i>F^{eq}_i$, for $i=A,B$. As an explicit example in the following we study the PKLSZDK game, i.e., the game $\mathcal{G}(1/2,1)$. For this game it becomes that $F_A+F_B=\frac{3}{4}\left(1+\frac{1}{4}\mathbb{B}\right)$, i.e., the sum of Alice's and Bob's average payoffs in classical, quantum, and no-signaling theories can be at most $9/8$, $3/4(1+1/\sqrt{2})$, and $3/2$, respectively.

\emph{(a) Fair strategy}: For fair strategy the average equilibrium payoffs of Alice and Bob are identical, which gives the condition, $2(m_0+n_0)+m_1+n_1=3$ and both Alice's and Bob's payoffs become $3/8(1+1/4\mathbb{B})$. Thus any no-signaling advice will be advantageous over the best fair classical strategy if,
\begin{equation}
\frac{3}{8}\left(1+\frac{1}{4}\mathbb{B}\right)>\frac{9}{16}~~\Rightarrow~~\mathbb{B}>2.\label{c2}
\end{equation}  
As described in \cite{Pappa15}, a quantum strategy is specified by the triple $\{\rho_{AB}, (\mathcal{M}^0_A,\mathcal{M}^1_A), (\mathcal{M}^0_B,\mathcal{M}^1_B)\}$ \cite{Pappa15}, where $\rho_{AB}$ is some bipartite quantum state provided to Alice and Bob as advice and $\mathcal{M}^k_i$ are two outcomes positive-operator-valued-measures (POVM) for $k\in\{0,1\}$, $i\in\{A,B\}$. For example, two qubit Werner class of states $\mathcal{W}_p=p|\psi^-\rangle\langle\psi^-|+(1-p)\frac{\mathbb{I}}{2}\otimes\frac{\mathbb{I}}{2}$, where $|\psi^-\rangle=\frac{1}{\sqrt{2}}(|0\rangle\otimes|1\rangle-|1\rangle\otimes|0\rangle)$ is the singlet state and $\mathbb{I}$ be the identity operator, satisfies the first condition for projective measurements performed by Alice and Bob. And for suitably chosen measurement they satisfy the second condition whenever $p>\frac{1}{\sqrt{2}}$. Moreover the optimal Bell-CHSH violation in quantum theory, i.e., $2\sqrt{2}$ is uniquely (up to local unitary freedom) achieved by singlet state with suitably chosen measurements \cite{Horodecki96}. This implies that the singlet state with corresponding measurement settings describe a fair equilibrium strategy in quantum theory. It is noteworthy that more general no-signaling advices can outperform best quantum strategy.  

\emph{(a) Unfair strategy}: From Table-\ref{table2} note that $(g_A^1,g_B^3)$ is an unfair (to Bob) classical equilibrium strategy with Alice's and Bob's payoffs $11/16$ and $7/16$, respectively for the game $\mathcal{G}(1/2,1)$. So a no-signaling advice will be considered advantageous over this strategy if $F_A>11/16$ and $F_B>7/16$ or equivalently to say:
\begin{eqnarray}
\frac{3}{2}\mathbb{B}+2(m_0+n_0)+(m_1+n_1)>8,
~\mbox{and}\label{c3}\\
\frac{3}{2}\mathbb{B}-2(m_0+n_0)-(m_1+n_1)>-2.\label{c4}
\end{eqnarray}
From conditions (\ref{c3}) and (\ref{c4}) it is clear that the no-signaling correlations need to be nonlocal to provide advantage over the classical unfair equilibrium strategy. If we consider the following one parameter family of no-signaling correlations:  
\begin{equation}
\mathcal{A}=q\mathcal{PR}+(1-q)\mathcal{D},\label{nl}
\end{equation}
where $0\le q\le1$ and $\mathcal{PR}=\{P_{\mathcal{PR}}(ab|ij)\}$ and $\mathcal{D}=\{P_{\mathcal{D}}(ab|ij)$ with, $P_{\mathcal{PR}}(ab|ij)=1/2~\mbox{if}~\alpha\oplus \beta=ij$ and $0$ otherwise, and $P_{\mathcal{D}}(ab|ij)=1~\mbox{if}~\alpha=\beta=1$ and $0$ otherwise,
then straightforward calculation gives,
\begin{equation}
F_A=\frac{11}{16}+\frac{q}{16},~~~~F_B=\frac{7}{16}+\frac{5q}{16},
\end{equation}
establishing advantage over the unfair classical equilibrium strategy for nonzero value of $q$. However this example does not solve the question whether quantum theory provides advantage over the classical strategies. This is because in Ref.\cite{Allcock09} it has been shown that in the asymptotic limit the correlation $\mathcal{A}$ of Eq.(\ref{nl}) can be distilled to maximally nonlocal $\mathcal{PR}$ correlation and hence quantum realization of this correlation is not possible. 

To find a quantum strategy better than classical unfair equilibrium strategy we first consider the singlet state as the advice. For the sharp as well as unsharp \cite{Unsharp} measurements performed by Alice or Bob on their respective parts the marginal probabilities are completely random and hence it is not possible to fulfill the condition (\ref{c3}) and (\ref{c4}) simultaneously. Though more general two outcome POVM can give biased marginal but still they are not useful (see \cite{Supply} for detail). So singlet state is no good in this purpose and similar argument applies for any bipartite state with completely mixed marginals. Naturally we then consider a pure state $|\psi\rangle_{AB}=a|00\rangle+b|11\rangle$ as the advice, with Alice's and Bob's measurement directions specified as, $M^k_i\equiv\left(\sin~\theta^k_i \cos~\phi^k_i,\sin~\theta^k_i \sin~\phi^k_i,\cos~\theta^k_i\right)$,
where $k\in\{0,1\}$ and $i\in\{A,B\}$. Interestingly, we find that such advice can fulfill the requirement. For example, if we take an advice with $a=0.9$ and the measurement directions $\theta^0_A=-\theta^0_B=-\frac{\pi}{15}$,  $\phi^0_A=-\phi^0_B=\frac{\pi}{2}$,
$\theta^1_i=\frac{\pi}{3}$, and $\phi^1_i=\frac{\pi}{2}$, then their average payoffs become $F_A=0.7066 (>\frac{11}{16})$ and $F_B=0.5163  (>\frac{7}{16})$ (see \cite{Supply}). However, this strategy is not a quantum equilibrium strategy. Because, given the same advice and same measurement settings for Alice, Bob can fix his measurement directions as  $\theta^{0(new)}_B=0.451517,~\theta^{1(new)}_B=1.25911,
~\phi^{0(new)}_B=-1.5708,~\phi^{1(new)}_B=1.5708$ so that his new payoff becomes $F_B^{new}=0.5213>F_B>\frac{7}{16}$. But for these new measurement choices Alice's payoff modifies to $\frac{11}{16}<F_A^{new}=0.6981<0.7066=F_A$. Even if the parties consider two outcome POVMs, the equilibrium will not be achieved.

At this point one can think of a stronger refinement of the equilibrium concept, known as social optimality \citep{Binmore98}. It is a choice of strategies, one by each player, that maximizes the sum of the players' payoffs. Of course, maximization of the sum of players' payoff may not necessarily lead to the satisfaction of all the participating players. Given the advice, $F_A+F_B=\frac{3}{4}(1+\frac{1}{2}\sqrt{1+4a^2(1-a^2)})=1.2266$, and one can find a strategy (see \cite{Supply}) such that $F^*_A=0.6978>\frac{11}{16}$ and $F^*_B=0.5288>\frac{7}{16}$. Clearly this $`*'$ strategy is a social optimal one, while the previous two are not. It is quite mentioning that quantum strategies also outperform the other classical unfair (unfair to Alice) equilibrium strategy. The analysis remains same if a mixed entangled advice is given instead of pure one. 

\emph{Discussions}: Nonlocality, arguably one of the most controversial issues in quantum foundations, has acquired lots of research attention of information theorists during last two decades due to its practical usefulness in several device independent information theoretic protocols. Very recently interesting  connection of this peculiar resource has been established in another very important branch of study, namely Bayesian game theory which has several applications in economics, social and political science, and psychology. Nonlocal correlations have been shown to be advantageous in common interest Bayesian games and also in conflicting games over the classical fair strategy \cite{Brunner13,Pappa15,Games}. In this work we have shown that such nonlocal correlations can outperform the unfair classical equilibrium strategies of such games. Moreover, quantum advices can provide unfair social optimal strategies better than the classical one. To prove this advantages we have considered a two parametric class of two players games which can have both fair and unfair classical equilibria and also only unfair equilibria, depending on the different parameters values. Although we have considered a particular class but our analysis points out the effectiveness of nonlocal advice over any classical correlation. We also completely characterize the no-signaling advices providing advantages in these game over the fair and unfair classical equilibrium strategies.      
	
\emph{Acknowledgments}: The authors would like to thank Guruprasad Kar for many stimulating discussions and fruitful suggestions. Discussion with Ashutosh Rai during his visit to ISI, Kolkata is gratefully acknowledged. AM thanks Council of
Scientific and Industrial Research, India for financial
support through Senior Research Fellowship (Grant No.
09/093(0148)/2012-EMR-I).



\section*{APPENDIX}
\subsection{Singlet state is no good for unfair strategy}
As we have discussed in the manuscript, a strategy will be advantageous over classical unfair equilibrium one if,
\begin{eqnarray}\label{c5}
\frac{3}{2}\mathbb{B}+\mathbb{K}&>&8,~~\mbox{and}\\\label{c6}
\frac{3}{2}\mathbb{B}-\mathbb{K}&>&-2,
\end{eqnarray}
where $\mathbb{K}=2(m_0+n_0)+m_1+n_1$, with $m_i$ ($n_i$) being the marginal of $+$ outcome of the $i^{th}$ measurement of Alice (Bob). For singlet state the marginal state of each party is completely mixed state.
Projective measurement $M_{\hat{a}}$ and unsharp  measurement $M_{\hat{a}}^{(\lambda)}$ along some direction $\hat{a}$, is given by,
\begin{eqnarray}
M_{\hat{a}}&=&\{\frac{1}{2}(\mathbb{I}+\hat{a}.\vec{\sigma}),\frac{1}{2}(\mathbb{I}-\hat{a}.\vec{\sigma})\},\\
M_{\hat{a}}^{(\lambda)}&=&\{\frac{1}{2}(\mathbb{I}+\lambda\hat{a}.\vec{\sigma}),\frac{1}{2}(\mathbb{I}-\lambda\hat{a}.\vec{\sigma})\},
\end{eqnarray}
where $\frac{1}{2}(\mathbb{I}\pm\hat{n}.\vec{\sigma})$ are projectors corresponding to up and down eigenstates, and $\frac{1}{2}(\mathbb{I}\pm\lambda\hat{n}.\vec{\sigma})$ being the corresponding unsharp effect with $0<\lambda\le 1$. For such measurements performed on the each part of the singlet state the marginal statistics is completely random. Hence the value of $\mathbb{K}$ is always $3$. Since the value of $\mathbb{B}$ can be at most $2\sqrt{2}$, the left hand side of Eq.(\ref{c5}) can be at most $3(1+\sqrt{2})< 8$. So singlet state along with sharp or unsharp measurements is no good for the purpose.

However considering a more general two outcome qubit measurement one can get the value of $\mathbb{K}$ very close to $6$ and hence having a possibility of satisfying the condition (\ref{c5}). At the same time the setting must have satisfy the condition (\ref{c6}). In the following we study these possibilities. The most general two outcome qubit measurement (POVM) along $\hat{m}$ can be expressed as $\{E_{\hat{m}}(\alpha,\mu),\mathbb{I}-E_{\hat{m}}(\alpha,\mu)\}$ \cite{Unsharp}, where 
\begin{eqnarray}
E_{\hat{m}}(\alpha,\mu)=\frac{1}{2}(\alpha\mathbb{I}+\mu\hat{m}.\vec{\sigma}),~~\mbox{with}\nonumber\\\label{pp}
0<\alpha\le 2,~~\mbox{and}~0\le\mu\le\min\{\alpha,2-\alpha\}.\label{p1}
\end{eqnarray} 
\begin{figure}[t!]
\centering
\includegraphics[height=5cm,width=8cm]{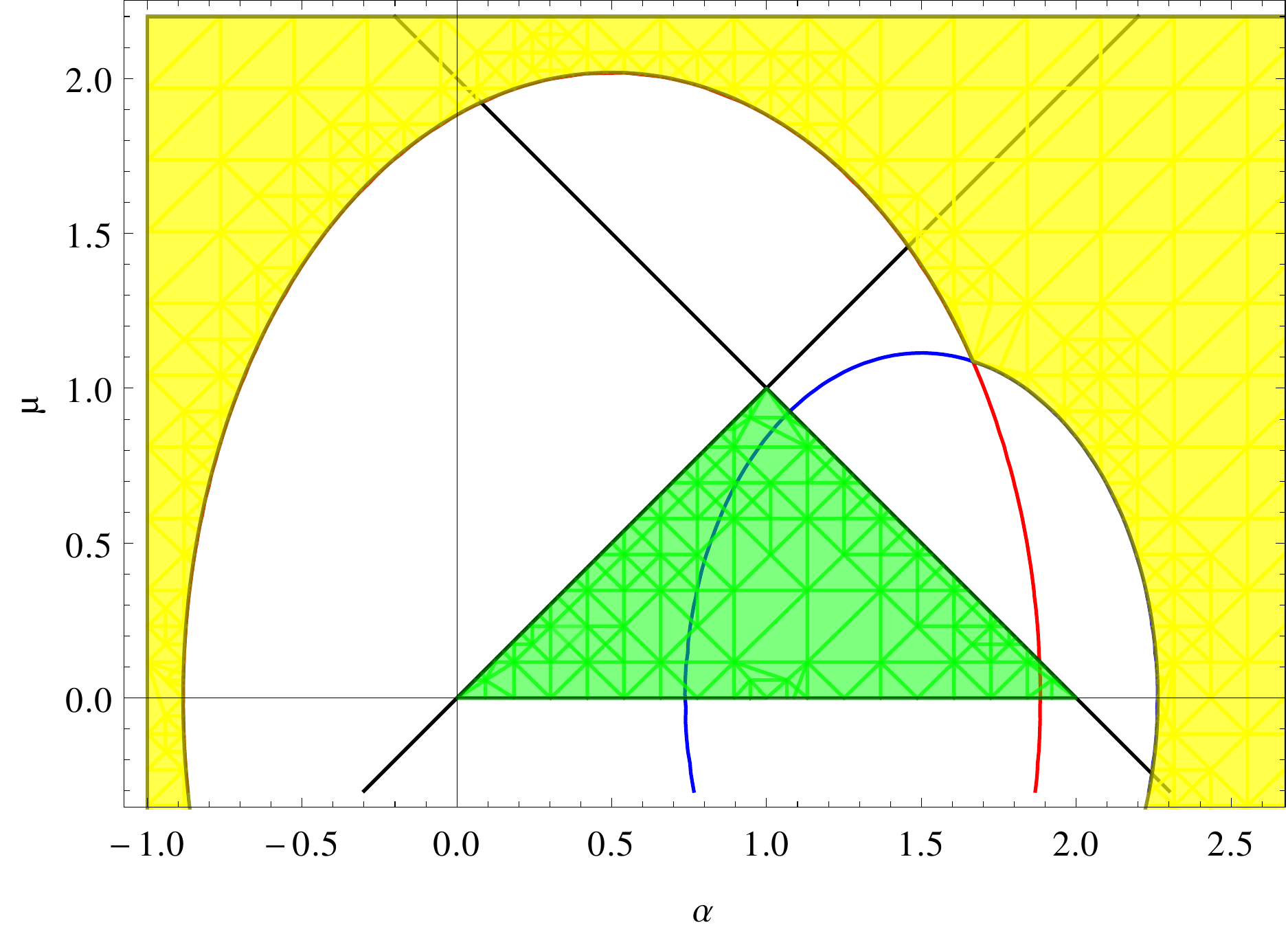}
	\caption{(Color on-line) (Color on-line) $\mathbb{B}_S=2.82$. Red curve is for (\ref{c7}) and blue is for (\ref{c8}). Allowed values of $(\alpha,\mu)$ satisfying condition (\ref{p1}) lie in the green region. Allowed values of $(\alpha,\mu)$ satisfying both the conditions (\ref{c7}) and (\ref{c8}) lies in the yellow regions.}\label{fig1}
\includegraphics[height=5cm,width=8cm]{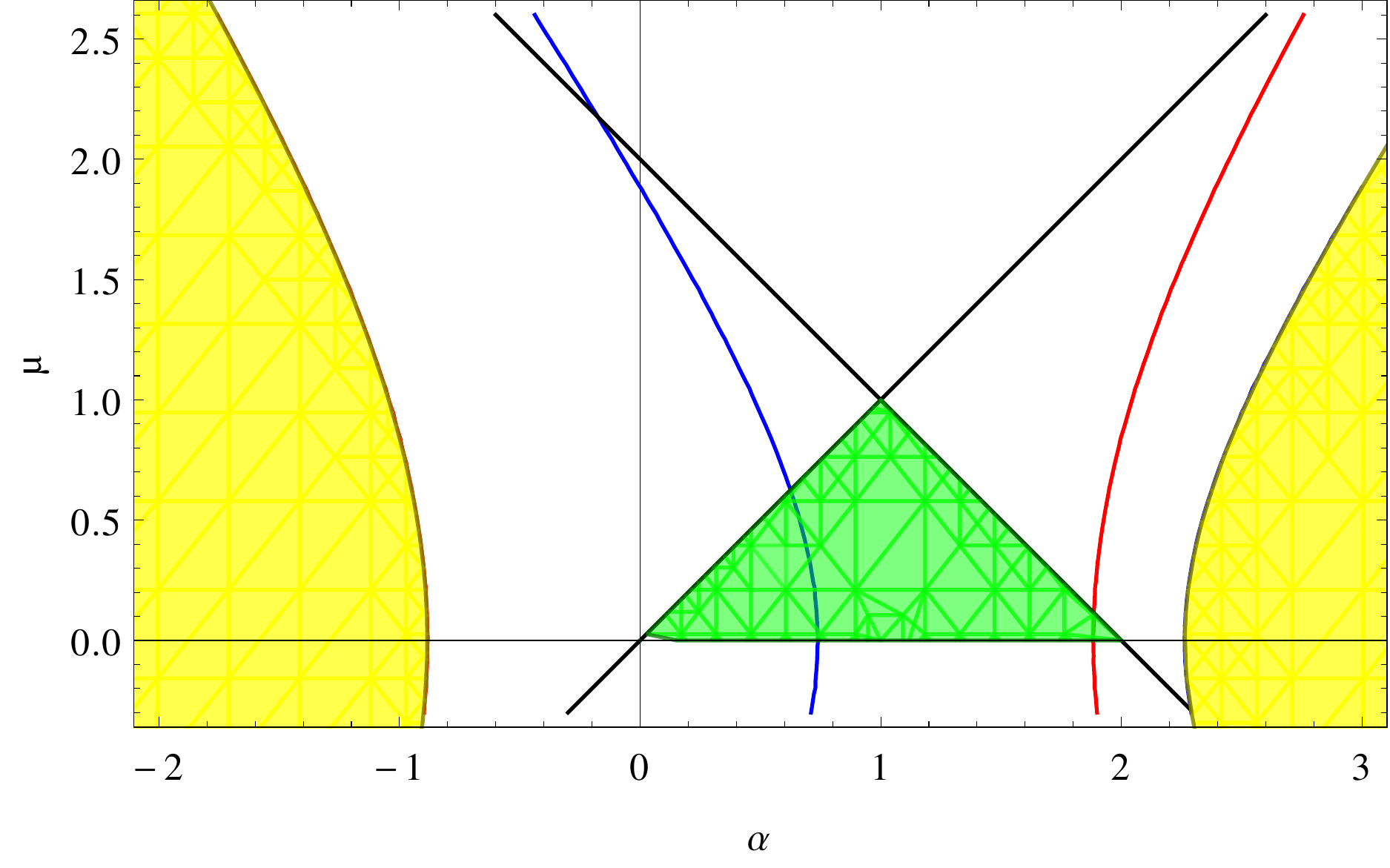}
	\caption{(Color on-line) $\mathbb{B}_S=-2.82$. Red curve is for (\ref{c7}) and blue is for (\ref{c8}). Allowed values of $(\alpha,\mu)$ satisfying condition (\ref{p1}) lie in the green region. Allowed values of $(\alpha,\mu)$ satisfying both the conditions (\ref{c7}) and (\ref{c8}) lies in the yellow regions.}\label{fig2}
\end{figure}  
If Alice performs the POVM $M\equiv\{E_{\hat{m}}(\alpha,\mu),\mathbb{I}-E_{\hat{m}}(\alpha,\mu)\}$ and Bob performs the POVM $N\equiv\{E_{\hat{n}}(\alpha,\mu),\mathbb{I}-E_{\hat{n}}(\alpha,\mu)\}$ on their respective part of the singlet state, the joint conditional probabilities read:
\begin{eqnarray}
P(++)&=&\frac{1}{4}[\alpha^2-\mu^2 \hat{m}.\hat{n}],\nonumber\\
P(+-)&=&\frac{1}{4}[\alpha(2-\alpha)+\mu^2 \hat{m}.\hat{n}],\nonumber\\
P(-+)&=&\frac{1}{4}[(2-\alpha)\alpha+\mu^2 \hat{m}.\hat{n}],\nonumber\\
P(--)&=&\frac{1}{4}[\alpha^2-\mu^2 \hat{m}.\hat{n}].
\end{eqnarray}
The marginal probabilities are therefore,
\begin{eqnarray}
P_{\hat{m}}(+)&:=&P(++)+P(+-)=\alpha/2,\nonumber\\
P_{\hat{n}}(+)&:=&P(-+)+P(--)=\alpha/2,\nonumber
\end{eqnarray}
and the expectation $\langle MN\rangle$ becomes,
\begin{eqnarray}
\langle MN\rangle &:=&P(++)-P(+-)-P(-+)+P(--)\nonumber\\
&=&(\alpha-1)^2-\mu^2\hat{m}.\hat{n}.\nonumber
\end{eqnarray}
The Bell-CHSH quantity and $\mathbb{K}$ are therefore,
\begin{eqnarray}
\mathbb{B}&=&2(\alpha-1)^2-\mu^2(\hat{m}_0.\hat{n}_0+\hat{m}_0.\hat{n}_1+\hat{m}_1.\hat{n}_0-\hat{m}_1.\hat{n}_1)\nonumber\\
&:=& 2(\alpha-1)^2+\mu^2\mathbb{B}_{S}, ~\mbox{and}\\
\mathbb{K}&=&3\alpha,
\end{eqnarray}
with $\mathbb{B}_S$ be the CHSH value for sharp measurement taking both $+$ve and $-$ve values. And the conditions (\ref{c5})-(\ref{c6}) look:
\begin{eqnarray}\label{c7}
\frac{(\alpha-1/2)^2}{23/12}+\frac{\mu^2}{23/6\mathbb{B}_S}&>&1,\\\label{c8}
\frac{(\alpha-3/2)^2}{7/12}+\frac{\mu^2}{7/6\mathbb{B}_S}&>&1.
\end{eqnarray}
With $\mathbb{B}_S$ taking $+$ve and $-$ve values, respectively, the equality of the conditions (\ref{c7})-(\ref{c8}) represent ellipses and hyperbolas. For $|\mathbb{B}_S|=2.82\simeq 2\sqrt{2}$ corresponding ellipses and hyperbolas are plotted in Fig.\ref{fig1} and Fig.\ref{fig2}, respectively. 
  
A choice of $(\alpha,\mu)$ will violate both the conditions (\ref{c7}) and (\ref{c8}) if it lies outside both the ellipse (hyperbolas), i.e., in the yellow region in Fig.\ref{fig1} (Fig.\ref{fig2}). However, 
the allowed values of $(\alpha,\mu)$ [compatible with condition (\ref{pp})] are in the green region, which has no overlap with the yellow region. Of course there are $(\alpha,\mu)$ outside one curve but inside other. Changing the value of $\mathbb{B}_S$, we find that there is no allowed $(\alpha,\mu)$ which can fulfill both the requirements imposed by the conditions (\ref{c7}) and (\ref{c8}). 
   
Therefore singlet state, even with most general two outcome POVM, is not an useful advice over the unfair classical equilibrium strategy in the PKLSZDK game. Similar kind of argument is also true for any two qubit state with both the marginal states completely mixed.    
   
\subsection{General strategy with non-maximally pure entangled advice} 
For an arbitrary pure non-maximally entangled state $|\psi\rangle_{AB} = a|00\rangle + b|11\rangle$ and the projective measurements performed by Alice and Bob for arbitrary directions, say,
\begin{eqnarray}
M^0_A&\equiv & (p_1, p_2, p_3);~~M^1_A\equiv (q_1, q_2,
q_3),\\
M^0_B&\equiv & (r_1, r_2, r_3);~~M^1_B\equiv (s_1, s_2, s_3),
\end{eqnarray}
the input-output probability distribution is listed in Table.\ref{table3}. Also one can express the measurement directions in polar coordinate, i.e.,
\begin{equation}
M^k_i\equiv\left(\sin~\theta^k_i \cos~\phi^k_i,\sin~\theta^k_i \sin~\phi^k_i,\cos~\theta^k_i\right).
\end{equation} 	 	 
\begin{widetext}
\begin{center}
\begin{table}[h!]
\caption{Input-output probability distribution for projective measurements performed by Alice and Bob on the state $|\psi\rangle_{AB}=a|00\rangle+b|11\rangle$.}
		\begin{tabular}{|p{1cm}||p{4cm}|p{4cm}|p{4cm}|p{4cm}|}
			\hline\hline
			$ $ & $~~~~~~~~~~~~~~~~++$ & $~~~~~~~~~~~~~~~~+-$ & $~~~~~~~~~~~~~~~~-+$ & $~~~~~~~~~~~~~~~~--$\\	
			\hline\hline
			$M^0_AM^0_B$ & $\frac{1}{4}\{2ab(r_1p_1-r_2p_2)+b^2(-1+r_3)(p_3-1)+a^2(r_3+1)(p_3+1)\}$ & $\frac{1}{4}\{2ab(-r_1p_1+r_2p_2)-b^2(1+r_3)(p_3-1)-a^2(r_3-1)(p_3+1)\}$ & $\frac{1}{4}\{2ab(-r_1p_1+r_2p_2)-a^2(1+r_3)(p_3-1)-b^2(r_3-1)(p_3+1)\}$ & $\frac{1}{4}\{2ab(r_1p_1-r_2p_2)+a^2(-1+r_3)(p_3-1)+b^2(r_3+1)(p_3+1)\}$ \\ \hline
			$M^0_AM^1_B$ & $\frac{1}{4}\{2ab(s_1p_1-s_2p_2)+b^2(-1+s_3)(p_3-1)+a^2(s_3+1)(p_3+1)\}$ & $\frac{1}{4}\{2ab(-s_1p_1+s_2p_2)-b^2(1+s_3)(p_3-1)-a^2(s_3-1)(p_3+1)\}$ & $\frac{1}{4}\{2ab(-s_1p_1+s_2p_2)-a^2(1+s_3)(p_3-1)-b^2(s_3-1)(p_3+1)\}$ & $\frac{1}{4}\{2ab(s_1p_1-s_2p_2)+a^2(-1+s_3)(p_3-1)+b^2(s_3+1)(p_3+1)\}$ \\ \hline
			$M^1_AM^0_B$ & $\frac{1}{4}\{2ab(r_1q_1-r_2q_2)+b^2(-1+r_3)(q_3-1)+a^2(r_3+1)(q_3+1)\}$ & $\frac{1}{4}\{2ab(-r_1q_1+r_2q_2)-b^2(1+r_3)(q_3-1)-a^2(r_3-1)(q_3+1)\}$ & $\frac{1}{4}\{2ab(-r_1q_1+r_2q_2)-a^2(1+r_3)(q_3-1)-b^2(r_3-1)(q_3+1)\}$ & $\frac{1}{4}\{2ab(r_1q_1-r_2q_2)+a^2(-1+r_3)(q_3-1)+b^2(r_3+1)(q_3+1)\}$ \\ \hline
			$M^1_AM^1_B$ & $\frac{1}{4}\{2ab(s_1q_1-s_2q_2)+b^2(-1+s_3)(q_3-1)+a^2(s_3+1)(q_3+1)\}$ & $\frac{1}{4}\{2ab(-s_1q_1+s_2q_2)-b^2(1+s_3)(q_3-1)-a^2(s_3-1)(q_3+1)\}$ & $\frac{1}{4}\{2ab(-s_1q_1+s_2q_2)-a^2(1+s_3)(q_3-1)-b^2(s_3-1)(q_3+1)\}$ & $\frac{1}{4}\{2ab(s_1q_1-s_2q_2)+a^2(-1+s_3)(q_3-1)+b^2(s_3+1)(q_3+1)\}$ \\ \hline 
		\end{tabular}\label{table3}
				\end{table}
	\end{center}
\end{widetext}
Given such a strategy $\{|\psi\rangle_{AB},(\mathcal{M}^0_A,\mathcal{M}^1_A),(\mathcal{M}^0_B,\mathcal{M}^1_b)\}$, where $\mathcal{M}^k_i$ be the projective measurement along $M^k_i$, the average payoffs of Alice and Bob can be calculated from the Table.\ref{table3}, using the expression of Eqs.(\ref{bell}), (\ref{fa}), and (\ref{fb}), and the expressions become:
\begin{eqnarray}
F_A&=&\frac{1}{8} \left[3+\frac{3}{2}ab(r_1(p_1+q_1)+s_1(p_1-q_1)-r_2(p_2+q_2)\right.\nonumber\\
&&~~~~~ -s_2(p_2-q_2))+\frac{3}{4}(r_3+s_3)(p_3-q_3)\nonumber\\
&&\left. ~~~~~~~+\frac{1}{4}(a^2-b^2)(q_3+s_3+2(p_3+r_3))  \right],   
\end{eqnarray}
\begin{eqnarray}
F_B&=&\frac{1}{8} \left[3+\frac{3}{2}ab(r_1(p_1+q_1)+s_1(p_1-q_1)-r_2(p_2+q_2)\right.\nonumber\\
&&~~~~~-s_2(p_2-q_2))+\frac{3}{4}(r_3+s_3)(p_3-q_3)\nonumber\\ 
&&\left. ~~~~~~~-\frac{1}{4}(a^2-b^2)(q_3+s_3+2(p_3+r_3))  \right].   
\end{eqnarray}	
The condition for fair strategies thus reads:
\begin{equation}
(a^2-b^2)(q_3+s_3+2(p_3+r_3)=0,
\end{equation}
which can be satisfied by non-maximally pure entangled state (i.e. $a\neq b$) with appropriately chosen measurement settings (i.e. $q_3+s_3+2p_3+2r_3=0$). However, here we are interested whether they can be useful for unfair strategies. Given the above expressions for $F_A$ and $F_B$ the aim is to search for strategies surpassing the classical unfair equilibrium one. And it can be done efficiently by using linear programming.  

\emph{Examples given in manuscript}: For an arbitrary $|\psi\rangle_{AB}=a|00\rangle+b|11\rangle$, if Alice and Bob choose their measurement directions as $\theta^0_i=-\frac{\pi}{15}$ , $\phi^0_i=\frac{\pi}{2}$, $\theta^1_i=\frac{\pi}{3}$ , $\phi^1_i=\frac{\pi}{2}$ then their average payoffs look:  
\begin{eqnarray}
 F_A &=&\frac{1}{8}\left[ \left(2a^2-\frac{1}{4}\right)\cos~\left(\frac{\pi}{15}\right)+\frac{3}{4}\cos^2~\left(\frac{\pi}{15}\right)\right.\nonumber\\
 &&\left.+\frac{3a\sqrt{3(1-a^2)}}{2}\sin~\left(\frac{\pi}{15}\right)-\frac{3a\sqrt{1-a^2}}{2}\sin^2\left(\frac{\pi}{15}\right)\right.\nonumber\\ 
&&\left.+\frac{45}{16}+\frac{9a\sqrt{1-a^2}}{8}+\frac{2a^2-1}{4}\right],
\end{eqnarray}
\begin{eqnarray}
F_B&=&\frac{1}{8}\left[ \left(\frac{7}{4}-2a^2\right)\cos~\left(\frac{\pi}{15}\right)+\frac{3}{4}\cos^2~\left(\frac{\pi}{15}\right)\right.\nonumber\\
&&\left.+\frac{3a\sqrt{3(1-a^2)}}{2}\sin~\left(\frac{\pi}{15}\right)
	 	-\frac{3a\sqrt{1-a^2}}{2}\sin^2\left(\frac{\pi}{15}\right)\right.\nonumber\\ 
&&\left.+\frac{45}{16}+\frac{9a\sqrt{1-a^2}}{8}-\frac{2a^2-1}{4}\right].
\end{eqnarray}
For $a=0.9$, we get $F_A=0.7066 (>\frac{11}{16})$ and $F_B=0.5163  (>\frac{7}{16})$. 
\begin{figure}[t!]
\centering
\includegraphics[height=5cm,width=8cm]{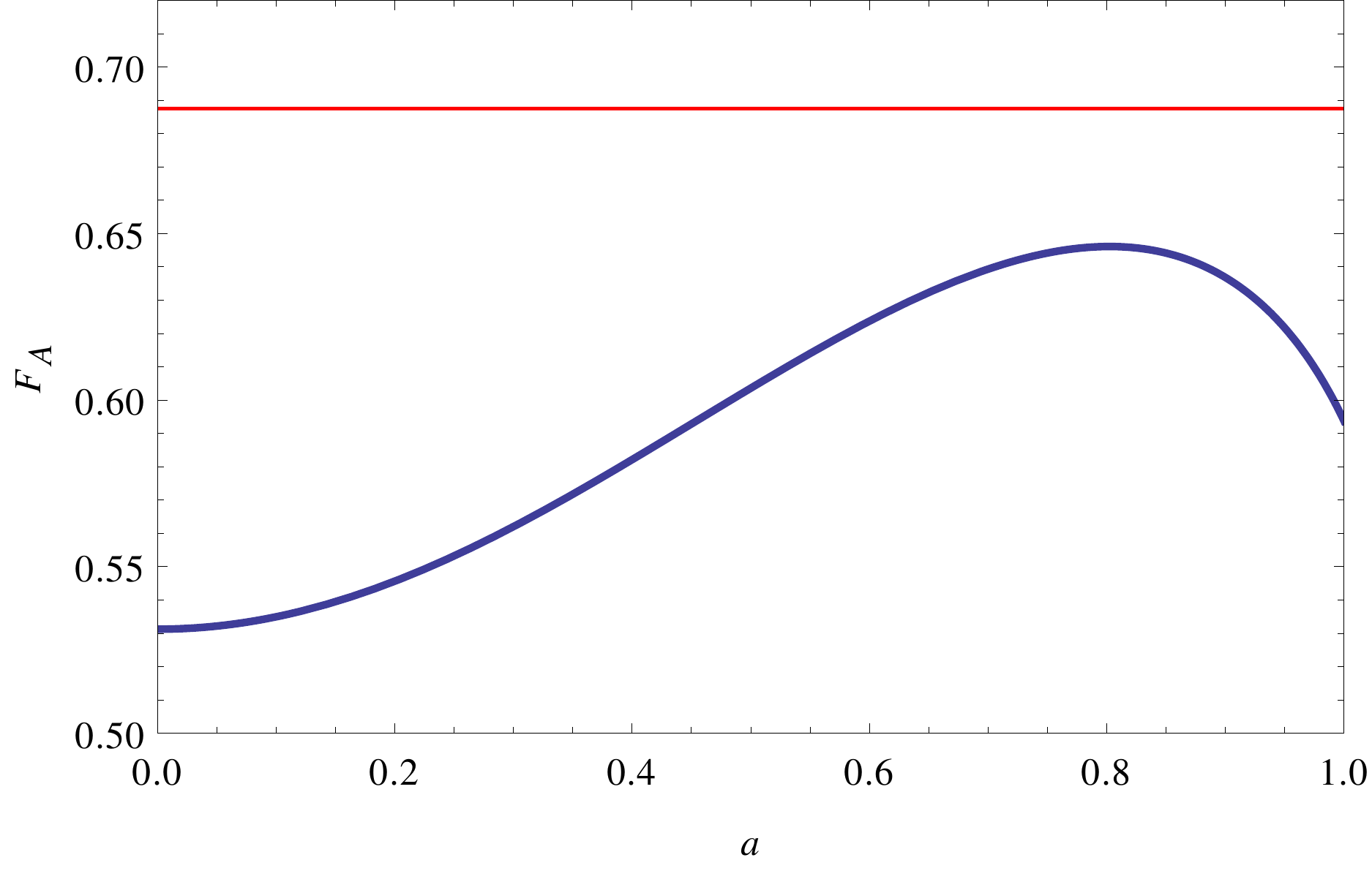}
	\caption{(Color on-line) Blue curve denotes the value of $F_A$ [Eq.(\ref{ng1})] and the red line is the value $11/16$. For every $a\in(0,1)$, $F_A$ is less than $11/16$.}\label{fig3}
\end{figure}

\emph{Social optimality strategy}: For the following measurement settings of Alice and Bob:
\begin{eqnarray}
\theta^0_A=0,~\theta^1_A\cong-1.5708,\nonumber\\
\phi^0_A\cong-2.3636,~\phi^1_A\cong0.777996,
\end{eqnarray}
\begin{eqnarray}
\theta^0_B\cong-0.6653,~\theta^1_B\cong0.6653,\nonumber\\
\phi^0_B\cong-0.7780,~\phi^1_B\cong-0.7780,
\end{eqnarray}
we get,$F_A\cong0.6978,~F_B\cong0.5288$, giving a social optimal strategy. At this point please note that, not all measurement settings giving optimal $\mathbb{B}$ for a quantum advice will not give $F_A>\frac{11}{16}$ and $F_B>\frac{7}{16}$. As for example, for the state $|\psi\rangle_{AB}=a|01\rangle+b|10\rangle$ the measurements choices $M^0_A\equiv(0,0,1)$, $M^1_A\equiv(1,0,0)$, $M^0_B\equiv(\sin\beta,0,\cos\beta)$, and $M^1_B\equiv(\sin\beta',0,\cos\beta')$, with $\cos\beta=-\cos\beta'=1/\sqrt{1+4a^2b^2}$, give the optimal Bell violation $2\sqrt{1+4a^2b^2}$ \cite{Gisin91}. However for this settings we get,
\begin{eqnarray}\label{ng1}
F_A&=&\frac{1}{16}\left(\frac{7+22 a^2-24 a^4}{2\sqrt{1+4 a^2-4 a^4}}+2a^2+5\right),\\\label{ng2}
F_B&=&\frac{1}{16}\left(\frac{5+26 a^2-24a^4}{2\sqrt{1+4 a^2-4 a^4}}-2a^2+7\right).
\end{eqnarray} 
For any value of $a\in(0,1)$, $F_A$ does not satisfy the required condition (see Fig.\ref{fig3}).


\begin{thebibliography}{99}
	
\bibitem{Bell64} J. S. Bell, ``On the Einstein Podolsky Rosen Paradox", Physics \emph{1} (3): 195–200 (1964), J. S. Bell, Speakable and Unspeakable in Quantum Mechanics (Cambridge University Press, 1987).

\bibitem{Reviw} N. Brunner, D. Cavalcanti, S. Pironio, V. Scarani, and S.Wehner, ``Bell nonlocality",
\href{http://journals.aps.org/rmp/abstract/10.1103/RevModPhys.86.419}{Rev. Mod. Phys. {\bf 86}, 839 (2014)}. 
	
\bibitem{Aspect81} A. Aspect, P. Grangier, and G. Roger, ``Experimental Tests of Realistic Local Theories via Bell's Theorem",\href{http://dx.doi.org/10.1103/PhysRevLett.47.460}{ Phys. Rev. Lett. {\bf 47}, 460 (1981)};
A. Aspect, P. Grangier, and G. Roger, ``Experimental Realization of Einstein-Podolsky-Rosen-Bohm Gedankenexperiment: A New Violation of Bell's Inequalities",  \href{http://dx.doi.org/10.1103/PhysRevLett.49.91}{Phys. Rev. Lett., {\bf 49}, 91 (1982)}.
	
\bibitem{loopfree} B. Hensen \emph{et al.} ``Loophole-free Bell inequality violation using electron spins separated by 1.3 kilometres", \href{http://www.nature.com/nature/journal/v526/n7575/full/nature15759.html}{Nature {\bf 526}, 682–686 (2015)}; 
M. Giustina \emph{et al.} ``Significant-Loophole-Free Test of Bell's Theorem with Entangled Photons',
\href{https://journals.aps.org/prl/abstract/10.1103/PhysRevLett.115.250401}{Phys. Rev. Lett. {\bf 115}, 250401 (2015)}; 
Lynden K. Shalm {\emph et al.} ``"Strong Loophole-Free Test of Local Realism", 
\href{https://journals.aps.org/prl/abstract/10.1103/PhysRevLett.115.250402}{Phys. Rev. Lett. {\bf 115}, 250402 (2015)}.
	
\bibitem{Applications} J. Barrett, L. Hardy, and A. Kent, ``No signaling and quantum key distribution", \href{http://dx.doi.org/10.1103/PhysRevLett.95.010503}{Phys. Rev. Lett. {\bf 95}, 010503 (2005)};
A. Ac\'{i}n, N. Gisin, and L. Masanes, ``From Bells theorem to secure quantum key distribution", \href{http://dx.doi.org/10.1103/PhysRevLett.97.120405}{Phys. Rev. Lett. {\bf 97}, 120405 (2006)};
S. Pironio et. al., ``Random numbers certified by Bells theorem", \href{http://www.nature.com/nature/journal/v464/n7291/full/nature09008.html}{Nature {\bf 464}, 1021 (2010)};
R. Colbeck and R. Renner, ``Free randomness can be amplified", \href{http://www.nature.com/nphys/journal/v8/n6/full/nphys2300.html}{Nat. Phys.{\bf 8}, 450 (2012)};
N. Brunner, S. Pironio, A. Ac\'{i}n, N. Gisin, A. A. Methot, and	V. Scarani, ``Testing the dimension of Hilbert spaces", \href{http://dx.doi.org/10.1103/PhysRevLett.100.210503}{Phys. Rev. Lett. {\bf 100}, 210503 (2008)};
R. Gallego, N. Brunner, C. Hadley, and A. Ac\'{i}n, ``Device independent tests of classical and quantum dimensions",  \href{http://dx.doi.org/10.1103/PhysRevLett.105.230501}{Phys. Rev. Lett. {\bf 105}, 230501 (2010)};
S. Das, M. Banik, A. Rai, MD R. Gazi, and S.Kunkri, ``Hardy's nonlocality argument as a witness for postquantum correlations",
\href{https://journals.aps.org/pra/abstract/10.1103/PhysRevA.87.012112}{Phys. Rev. A {\bf 87}, 012112 (2013)};
A. Mukherjee, A. Roy, S. S. Bhattacharya, S. Das, Md. R. Gazi, and M. Banik, ``Hardy's test as a device-independent dimension witness",
\href{https://journals.aps.org/pra/abstract/10.1103/PhysRevA.92.022302}{Phys. Rev. A {\bf 92}, 022302 (2015)};
A. Chaturvedi and M. Banik, ``Measurement-device–independent randomness from local entangled states",
\href{http://iopscience.iop.org/article/10.1209/
0295-5075/112/30003/meta;jsessionid=D6C96ABB3E61
C42C542A9553E8A4F4DC.c3.iopscience.cld.iop.org}{EPL {\bf 112}, 30003 (2015)}.

\bibitem{Brunner13} N. Brunner and N. Linden, ``Connection between Bell nonlocality and Bayesian game theory", \href{http://www.nature.com/ncomms/2013/130703/ncomms3057/full/ncomms3057.html#references}{Nature Communications {\bf 4}, 2057 (2013)}.

\bibitem{Pappa15} A. Pappa \emph{et al.} ``Nonlocality and Conflicting Interest Games", 
\href{http://journals.aps.org/prl/abstract/10.1103/PhysRevLett.114.020401}{Phys. Rev. Lett. {\bf 114}, 020401 (2015)}.

\bibitem{Clauser69} J. F. Clauser, M. A. Horne, A. Shimony and R. A. Holt, ``Proposed experiment to test local hidden-variable theories", \href{http://dx.doi.org/10.1103/PhysRevLett.23.880}{Phys. Rev. Lett. {\bf 23}, 880 (1969)}.

\bibitem{Harsanyi67} J. C. Harsanyi, ``Games with incomplete information played by Bayesian players", \href{http://pubsonline.informs.org/doi/abs/10.1287/mnsc.14.3.159}{Manage. Sci. {\bf 14}, 159–183 (Part I)}; \href{http://pubsonline.informs.org/doi/abs/10.1287/mnsc.14.5.320}{{\bf 14} (5) 320-334 (Part II)}; \href{http://pubsonline.informs.org/doi/abs/10.1287/mnsc.14.7.486}{{\bf 14} (7): 486-502 (Part III), (1967)}.

\bibitem{Osborne03}  M. J. Osborne, An Introduction to Game Theory (Oxford University Press, New York, 2003).

\bibitem{Aumann74} R. J. Aumann, ``Subjectivity and correlation in randomized strategies",
\href{http://www.sciencedirect.com/science/article/pii/0304406874900378}{Journal of mathematical economics {\bf 1}, 67 (1974)}.

\bibitem{Greenberger90} D. M. Greenberger, M. A. Horne, A. Shimony and A. Zeilinger,  ``Bell’s theorem without inequalities", \href{http://scitation.aip.org/content/aapt/journal/ajp/58/12/10.1119/1.16243}{Am. J. Phys. {\bf 58}, 1131 (1990)}.
	
\bibitem{Mermin90a} N. D. Mermin, ``Quantum Mysteries Revisited", \href{http://scitation.aip.org/content/aapt/journal/ajp/58/8/10.1119/1.16503}{Am. J. Phys. {\bf 58}, 731 (1990)}

\bibitem{Mermin90b} N. D. Mermin, ``Simple unified form for the major no-hidden-variables theorems", \href{http://journals.aps.org/prl/abstract/10.1103/PhysRevLett.65.3373}{Phys. Rev. Lett. 65, 3373 (1990)}

\bibitem{Peres90} A. Peres, ``Incompatible results of quantum measurements", \href{http://www.sciencedirect.com/science/article/pii/037596019090172K}{Physics Letters A, {\bf 151}, 107 (1990)}

\bibitem{Kerenidis04} Z. Bar-Jossef, T. S. Jayram, and I. Kerenidis, in Proceedings of the 36th Annual ACM Symposium on Theory of Computing (ACM, New York, 2004), pp. 128 –137
	
\bibitem{Buhrman11} H. Buhrman, O. Regev, G. Scarpa, and R. de Wolf, in Proceedings of the 26th IEEE Annual Conference on Computational Complexity (IEEE Computer Society, Washington, DC, 2011), pp. 157 –166.

\bibitem{Papadimitriou08} C H. Papadimitriou and T. Roughgarden, ``Computing Correlated Equilibria in Multi-Player Games", J. ACM {\bf 55}(3), 1-29 (2008).
	
\bibitem{Gazi13} Md. R. Gazi, M. Banik, S. Das, A. Rai, and S. Kunkri, ``Macroscopic locality with equal bias reproduces with high fidelity a quantum distribution achieving the Tsirelson's bound",
\href{https://journals.aps.org/pra/abstract/10.1103/PhysRevA.88.052115}{Phys. Rev. A {\bf 88}, 052115 (2013)}. 

\bibitem{Fine82} A. Fine, ``Hidden Variables, Joint Probability, and the Bell Inequalities"
\href{http://journals.aps.org/prl/abstract/10.1103/PhysRevLett.48.291}{Phys. Rev. Lett. {\bf 48}, 291 (1982)}.

\bibitem{Barrett05} J. Barrett, N. Linden, S. Massar, S. Pironio, S. Popescu, and D. Roberts, ``Nonlocal correlations as an information-theoretic resource",
\href{http://journals.aps.org/pra/abstract/10.1103/PhysRevA.71.022101}{Phys. Rev. A {\bf 71}, 022101 (2005)}.

\bibitem{Tsirelson80} B. S. Tsirelson, ``Quantum generalizations of Bell’s inequality", \href{http://link.springer.com/article/10.1007\%2FBF00417500}{Lett. Math. Phys. {\bf 4}, 93 (1980)}.

\bibitem{Popescu94} S. Popescu, D. Rohrlich, ``Quantum nonlocality as an axiom", \href{http://link.springer.com/article/10.1007\%2FBF02058098}{Found. Phys. {\bf 24}, 379 (1994)}.

\bibitem{Horodecki96} R. Horodecki and P. Horodecki, ``Perfect correlations in the Einstein-Podolsky-Rosen experiment and Bell's inequalities"
\href{http://www.sciencedirect.com/science/article/pii/0375960195009051}{Phys. Lett. A {\bf 201}, 227 (1996)}.

\bibitem{Allcock09} J. Allcock \emph{et al.} ``Closed sets of nonlocal correlations",
\href{http://journals.aps.org/pra/abstract/10.1103/PhysRevA.80.062107}{Phys. Rev. A {\bf 80}, 062107 (2009)}.

\bibitem{Unsharp} P. Busch, ``Unsharp reality and joint measurements for spin observables",
\href{https://journals.aps.org/prd/abstract/10.1103/PhysRevD.33.2253}{Phys. Rev. D {\bf 33}, 2253 (1986)}; 
A. Rai, MD. R. Gazi, M. Banik, S. Das, and S. Kunkri, ``Local simulation of singlet statistics for a restricted set of measurements",
\href{http://iopscience.iop.org/article/10.1088/1751-8113/45/47/475302/meta;jsessionid=BFCE06020FD3DE3E4E23833DC2D1A70A.c1.iopscience.cld.iop.org}{J. Phys. A: Math. Theor. {\bf 45}, 475302 (2012)}.

\bibitem{Supply} See the appendix.

\bibitem{Binmore98} K. Binmore, ``Just Playing: Game Theory and the Social Contract II", MIT Press, (ISBN 0-262-02444-6), (1998).

\bibitem{Gisin91} N. Gisin, ``Bell's inequality holds for all non-product states",
\href{http://www.sciencedirect.com/science/article/pii/037596019190805I}{Phys. Lett. A {\bf 154}, 201 (1991)}.

\bibitem{Games} A. N. Duman, ``Game Theory in Categorical Quantum Mechanics",
\href{http://arxiv.org/abs/1405.4428}{arXiv:1405.4428}; 
A. Iqbal, J. M. Chappell, D. Abbott, ``Social optimality in quantum Bayesian games", Physica A: Stat. Mech. Appli. {\bf 436}, 798 (2015);
R. Alonso-Sanz, ``A cellular automaton implementation of a quantum battle of the sexes game with imperfect information",
\href{http://link.springer.com/article/10.1007/
s11128-015-1080-3}{Quan. Inf. Process. {\bf 14}, 3639 (2015)}.
 
\end{thebibliography}
\end{document}